\documentclass[12pt,a4paper]{amsart}
\usepackage{amsfonts}
\usepackage[latin1]{inputenc}
\usepackage{amsmath,amssymb,amsthm,bm}
\usepackage{graphics}
\usepackage{epsfig,rotating}

\theoremstyle{definition}

\def\<{\langle}
\def\>{\rangle}

\input{tcilatex}

\begin{document}
\title{What are the systems that decohere?}
\author{Mario Castagnino}
\address{CONICET-IAFE-IFIR-Universidad de Buenos Aires}
\author{Sebastian Fortin}
\address{CONICET-IAFE-Universidad de Buenos Aires}
\author{Olimpia Lombardi}
\address{CONICET-Universidad de Buenos Aires}

\begin{abstract}
The fact that the Environment Induced Decoherence approach offers no general
criterion to decide where to place the \textquotedblleft
cut\textquotedblright\ between system and\ environment has been considered
as a serious conceptual problem of the proposal. \ In this letter we argue
that this is actually a pseudo-problem, which is dissolved by the fact that
decoherence is a phenomenon relative to the relevant observables selected by
the measuring arrangement. We also show that, when the spin-bath model is
studied from this perspective, certain unexpected results are obtained, as
that of a system decohering in interaction with a very small environment.
\end{abstract}

\maketitle

\paragraph{\textbf{Introduction.}}

Environment Induced Decoherence (EID), which turns the coherent state of an
open system into a decohered mixture, is the clue for the account of the
emergence of classicality from quantum mechanics \cite{Paz-Zurek}, \cite%
{Zurek-2003}.\ Therefore, the split of the universe into the system $S$ and
the environment $E$ is essential for EID. \ However, since the environment
may be \textquotedblleft external\textquotedblright\ or \textquotedblleft
internal\textquotedblright , the EID approach offers no general criterion to
decide where to place the \textquotedblleft cut\textquotedblright\ between
system and\ environment. Zurek considers this fact as a problem for his
proposal: \textquotedblleft \textit{In particular, one issue which has been
often taken for granted is looming big, as a foundation of the whole
decoherence program. It is the question of what are the `systems' which play
such a crucial role in all the discussions of the emergent classicality.}%
\textquotedblright\ (\cite{Zurek-cut}). The aim of this letter is to argue
that such a \textquotedblleft looming big\textquotedblright\ problem is
actually a pseudo-problem, which is dissolved by the fact that decoherence
is a phenomenon relative to the relevant observables selected in each
particular case. Precisely, if $\mathcal{O}$ is the space of all the
observables of a closed system, $\mathcal{O}_{R}\subset \mathcal{O}$ is the
space of the \textit{relevant} observables, that is, those that can be
experimentally measured. Since decoherence depends on the space $\mathcal{O}%
_{R}$ \ considered, and $\mathcal{O}_{R}$ changes with the change of the
measuring arrangement, decoherence turns out to be a phenomenon \textit{%
relative} to that arrangement.

Let us stress that we use the word `relative'\ strictly with the same
meaning as in special relativity, where it has no subjective content: a
reference frame is defined by a set of clocks and rules at rest in an
inertial system, and this set is the measuring arrangement. Analogously, a
quantum measuring arrangement is a set of devices having experimental access
only to the observables $O_{R}\in \mathcal{O}_{R}$; so, it is that
arrangement what defines, relatively, the system and its environment. With a
certain arrangement, the physicist may observe the decoherence of the system
so defined and the emergence of classicality in that system. But a different
arrangement defines a different system which may not decohere and, as a
consequence, retains its quantum behavior.

We will develop our argument by analyzing the well-known spin-bath model
from the general theoretical framework for decoherence presented in a
previous work \cite{Gen-fram}.

\paragraph{\textbf{The spin-bath model.}}

The spin-bath model is a very simple model that has been exactly solved in
previous papers (see \cite{Zurek-1982}). Let us consider a closed system $%
U=P+P_{i}$ where (i) $P$ is a spin-1/2 particle represented in the Hilbert
space $\mathcal{H}_{P}$, and (ii) the $P_{i}$ are $N$ spin-1/2 particles,
each one of which is represented in its own Hilbert space $\mathcal{H}_{i}$.
The complete Hilbert space of the composite system $U$\ is, $\mathcal{H}=%
\mathcal{H}_{P}\bigotimes\limits_{i=1}^{N}\mathcal{H}_{i}$. In the particle $%
P$, the two eigenstates of the spin operator $S_{S,\overrightarrow{v}}$\ in
direction $\overrightarrow{v}$ are $\left\vert \Uparrow \right\rangle $ and $%
\left\vert \Downarrow \right\rangle $, such that $S_{S,\overrightarrow{v}%
}\left\vert \Uparrow \right\rangle =\frac{1}{2}\left\vert \Uparrow
\right\rangle $\ and\ $S_{S,\overrightarrow{v}}\left\vert \Downarrow
\right\rangle =-\frac{1}{2}\left\vert \Downarrow \right\rangle $. In each
particle $P_{i}$, the two eigenstates of the corresponding spin operator $%
S_{i,\overrightarrow{v}}$\ in direction $\overrightarrow{v}$ are $\left\vert
\uparrow _{i}\right\rangle $ and $\left\vert \downarrow _{i}\right\rangle $,
such that $S_{i,\overrightarrow{v}}\left\vert \uparrow _{i}\right\rangle =%
\frac{1}{2}\left\vert \uparrow _{i}\right\rangle $ and\ $S_{i,%
\overrightarrow{v}}\left\vert \downarrow _{i}\right\rangle =\frac{1}{2}%
\left\vert \downarrow _{i}\right\rangle $. Therefore, a pure initial state
of $U$ reads%
\begin{equation}
|\psi _{0}\rangle =(a\left\vert \Uparrow \right\rangle +b\left\vert
\Downarrow \right\rangle )\bigotimes_{i=1}^{N}(\alpha _{i}|\uparrow
_{i}\rangle +\beta _{i}|\downarrow _{i}\rangle )  \label{3.3}
\end{equation}%
where the coefficients $a$, $b$, $\alpha _{i}$, $\beta _{i}$ are such that
satisfy $\left\vert a\right\vert ^{2}+\left\vert b\right\vert ^{2}=1$ and $%
\left\vert \alpha _{i}\right\vert ^{2}+\left\vert \beta _{i}\right\vert
^{2}=1$. Usually these numbers (and also the $g_{i}$ below) are taken as
aleatory numbers. If $P$ interacts with each one of the $P_{i}$ but the $%
P_{i}$ do not interact with each other, the total Hamiltonian $H$ of the
composite system $U$ results (see \cite{Zurek-1982}, \cite{Max})%
\begin{equation}
H=H_{SE}=S_{S,\overrightarrow{v}}\otimes \sum_{i=1}^{N}2g_{i}S_{i,%
\overrightarrow{v}}\bigotimes_{j\neq i}^{N}I_{j}  \label{3.4}
\end{equation}%
where $I_{j}$ is the identity operator on the subspace $\mathcal{H}_{j}$, $%
S_{S,\overrightarrow{v}}=\frac{1}{2}\left( \left\vert \Uparrow \right\rangle
\left\langle \Uparrow \right\vert -\left\vert \Downarrow \right\rangle
\left\langle \Downarrow \right\vert \right) $ and $S_{i,\overrightarrow{v}}=%
\frac{1}{2}\left( \left\vert \uparrow _{i}\right\rangle \left\langle
\uparrow _{i}\right\vert -\left\vert \downarrow _{i}\right\rangle
\left\langle \downarrow _{i}\right\vert \right) $. Under the action of $H$,
the state $|\psi _{0}\rangle $ evolves as $\left\vert \psi (t)\right\rangle
=a\left\vert \Uparrow \right\rangle |\mathcal{E}_{\Uparrow }(t)\rangle
+b\left\vert \Downarrow \right\rangle |\mathcal{E}_{\Downarrow }(t)\rangle $
where $\left\vert \mathcal{E}_{\Uparrow }(t)\right\rangle =\left\vert 
\mathcal{E}_{\Downarrow }(-t)\right\rangle $ and%
\begin{equation}
\left\vert \mathcal{E}_{\Uparrow }(t)\right\rangle
=\bigotimes_{i=1}^{N}\left( \alpha _{i}\,e^{ig_{i}t/2}\,\left\vert \uparrow
_{i}\right\rangle +\beta _{i}\,e^{-ig_{i}t/2}\,\left\vert \downarrow
_{i}\right\rangle \right)  \label{3.6}
\end{equation}

If $\mathcal{O}$ is the space of observables of the whole system $U$, let us
consider a space of relevant observables $\mathcal{O}_{R}\subset \mathcal{O}$
such that $O_{R}\in \mathcal{O}_{R}$ reads%
\begin{equation}
O_{R}=\left( 
\begin{array}{c}
s_{\Uparrow \Uparrow }\left\vert \Uparrow \right\rangle \left\langle
\Uparrow \right\vert \\ 
+s_{\Uparrow \Downarrow }\left\vert \Uparrow \right\rangle \left\langle
\Downarrow \right\vert \\ 
+s_{\Downarrow \Uparrow }\left\vert \Downarrow \right\rangle \left\langle
\Uparrow \right\vert \\ 
+s_{\Downarrow \Downarrow }\left\vert \Downarrow \right\rangle \left\langle
\Downarrow \right\vert%
\end{array}%
\right) \bigotimes_{i=1}^{N}\left( 
\begin{array}{c}
\epsilon _{\uparrow \uparrow }^{(i)}\left\vert \uparrow _{i}\right\rangle
\left\langle \uparrow _{i}\right\vert \\ 
+\epsilon _{\downarrow \downarrow }^{(i)}\left\vert \downarrow
_{i}\right\rangle \left\langle \downarrow _{i}\right\vert \\ 
+\epsilon _{\downarrow \uparrow }^{(i)}\left\vert \downarrow
_{i}\right\rangle \left\langle \uparrow _{i}\right\vert \\ 
+\epsilon _{\uparrow \downarrow }^{(i)}\left\vert \uparrow _{i}\right\rangle
\left\langle \downarrow _{i}\right\vert%
\end{array}%
\right)  \label{3.7}
\end{equation}%
Since the operators $O_{R}$ are Hermitian, the diagonal components $%
s_{\Uparrow \Uparrow }$, $s_{\Downarrow \Downarrow }$, $\epsilon _{\uparrow
\uparrow }^{(i)}$,$\epsilon _{\downarrow \downarrow }^{(i)}$ are real
numbers and the off-diagonal components are complex numbers satisfying $%
s_{\Uparrow \Downarrow }=s_{\Downarrow \Uparrow }^{\ast }$, $\epsilon
_{\uparrow \downarrow }^{(i)}=\epsilon _{\downarrow \uparrow }^{(i)\ast }$.
Then, the expectation value of the observable $O$ in the state $\left\vert
\psi (t)\right\rangle $ can be computed as%
\begin{eqnarray}
\langle O_{R}\rangle _{\psi (t)} &=&(|a|^{2}s_{\Uparrow \Uparrow
}+|b|^{2}s_{\Downarrow \Downarrow })\,\Gamma _{0}(t)  \notag \\
&&+2\func{Re}\,[ab^{\ast }\,s_{\Downarrow \Uparrow }\,\Gamma _{1}(t)]
\label{3.8}
\end{eqnarray}%
where (see eqs. (23) and (24) in \cite{Max}) 
\begin{align}
\Gamma _{0}(t)& =\prod_{i=1}^{N}\left[ 
\begin{array}{c}
|\alpha _{i}|^{2}\epsilon _{\uparrow \uparrow }^{(i)}+\alpha _{i}{}^{\ast
}\beta _{i}\epsilon _{\uparrow \downarrow }^{(i)}e^{-ig_{i}t} \\ 
+|\beta _{i}|^{2}\epsilon _{\downarrow \downarrow }^{(i)}+(\alpha
_{i}{}^{\ast }\beta _{i}\epsilon _{\uparrow \downarrow }^{(i)})^{\ast
}e^{ig_{i}t}%
\end{array}%
\right]  \label{3.9} \\
\Gamma _{1}(t)& =\prod_{i=1}^{N}\left[ 
\begin{array}{c}
|\alpha _{i}|^{2}\epsilon _{\uparrow \uparrow }^{(i)}e^{ig_{i}t}+|\beta
_{i}|^{2}\epsilon _{\downarrow \downarrow }^{(i)}e^{-ig_{i}t} \\ 
+\alpha _{i}{}^{\ast }\beta _{i}\epsilon _{\uparrow \downarrow
}^{(i)}+(\alpha _{i}{}^{\ast }\beta _{i}\epsilon _{\uparrow \downarrow
}^{(i)})^{\ast }%
\end{array}%
\right]  \label{3.10}
\end{align}

As a generalization of the usual presentations, we will study two different
ways of splitting the whole closed system $U$ into a relevant part and its
environment, by considering different choices for the space $\mathcal{O}_{R}$%
.

\paragraph{\textbf{Case 1: Observing the particle }$P$\textbf{.}}

In the typical situation studied by the EID approach, the system $S$ is
simply the particle $P$, ant the remaining particles $P_{i}$ are the
environment. Therefore, the relevant observables $O_{R}\in \mathcal{O}_{R}$
are those corresponding to $P$, and are obtained from eq. (\ref{3.7}) by
making $\epsilon _{\uparrow \uparrow }^{(i)}=\epsilon _{\downarrow
\downarrow }^{(i)}=1,$ $\epsilon _{\uparrow \downarrow }^{(i)}=0$:%
\begin{equation}
O_{R}=\left( \sum_{s,s^{\prime }=\Uparrow ,\Downarrow }s_{ss^{\prime
}}|s\rangle \langle s^{\prime }|\right)
\bigotimes_{i=1}^{N}I_{i}=O_{S}\bigotimes_{i=1}^{N}I_{i}  \label{3.11}
\end{equation}%
The expectation value of these observables is given by%
\begin{equation}
\langle O_{R}\rangle _{\psi (t)}=|a|^{2}\,s_{\Uparrow \Uparrow
}+|b|^{2}\,s_{\Downarrow \Downarrow }+2\func{Re}[ab^{\ast }\,s_{\Downarrow
\Uparrow }\,r_{1}(t)]  \label{3.12}
\end{equation}%
where 
\begin{equation}
r_{1}(t)=\prod_{i=1}^{N}\left[ |\alpha _{i}|^{2}e^{ig_{i}t}+|\beta
_{i}|^{2}e^{-ig_{i}t}\right]  \label{3.13}
\end{equation}%
By comparing eq. (\ref{3.12}) with eq. (\ref{3.8}), we see that in this case 
$\Gamma _{0}(t)=1$ and $\Gamma _{1}(t)=r_{1}(t)$. Moreover,

\begin{equation}
|r_{1}(t)|^{2}=\prod_{i=1}^{N}(|\alpha _{i}|^{4}+|\beta _{i}|^{4}+2|\alpha
_{i}|^{2}|\beta _{i}|^{2}\cos 2g_{i}t)  \label{3.14}
\end{equation}%
Since $|\alpha _{i}|^{2}+|\beta _{i}|^{2}=1$, then 
\begin{align}
& \max_{t}(|\alpha _{i}|^{4}+|\beta _{i}|^{4}+2|\alpha _{i}|^{2}|\beta
_{i}|^{2}\cos 2g_{i}t)  \notag \\
& =\left( \left( |\alpha _{i}|^{2}+|\beta _{i}|^{2}\right) ^{2}\right) =1
\label{3.15}
\end{align}%
and%
\begin{eqnarray}
&&\min_{t}\left( \left\vert \alpha _{i}\right\vert ^{4}+\left\vert \beta
_{i}\right\vert ^{4}+2\left\vert \alpha _{i}\right\vert ^{2}\left\vert \beta
_{i}\right\vert ^{2}\cos \left( 2g_{i}t\right) \right)  \notag \\
&=&\left( \left( |\alpha _{i}|^{2}-|\beta _{i}|^{2}\right) ^{2}\right)
=\left( 2\left\vert \alpha _{i}\right\vert ^{2}-1\right) ^{2}  \label{3.15'}
\end{eqnarray}%
If the coefficients $g_{i}$, $\alpha _{i}$ and $\beta _{i}$ are aleatory
numbers, then $(|\alpha _{i}|^{4}+|\beta _{i}|^{4}+2|\alpha _{i}|^{2}|\beta
_{i}|^{2}\cos 2g_{i}t)$ is an aleatory number which, if $t\neq 0$,
fluctuates between $1$ and $\left( 2\left\vert \alpha _{i}\right\vert
^{2}-1\right) ^{2}$. Let us note that, since the $\left\vert \alpha
_{i}\right\vert ^{2}$ and the $\left\vert \beta _{i}\right\vert ^{2}$ are
aleatory numbers in the closed interval $\left[ 0,1\right] $, when the
environment has many particles (that is, when $N\rightarrow \infty $), the
statistical value of the cases $\left\vert \alpha _{i}\right\vert ^{2}=1$, $%
\left\vert \beta _{i}\right\vert ^{2}=1$, $\left\vert \alpha _{i}\right\vert
^{2}=0$ and $\left\vert \beta _{i}\right\vert ^{2}=0$ is zero. In this case,
eq. (\ref{3.14}) for $|r_{1}(t)|^{2}$ is an infinite product of numbers
belonging to the open interval $\left( 0,1\right) $. \ As a consequence (see 
\cite{Paz-Zurek}, \cite{Zurek-2003}), 
\begin{equation}
\lim_{N\rightarrow \infty }r_{1}(t)=0  \label{3.16}
\end{equation}

In order to know the time-behavior of the expectation value of eq. (\ref%
{3.12}), we have to compute the time-behavior of $r_{1}(t)$. If we know that 
$r_{1}(0)=1$ for $N\rightarrow \infty $, and that $\lim_{N\rightarrow \infty
}r_{1}(t)=0$ for any $t\neq 0$, it can be expected that, for $N$ finite, $%
r_{1}(t)$ will evolve in time from $r_{1}(0)=1$ to a very small value. \
Moreover, $r_{1}(t)$ is a periodic function because it is a product of
periodic functions with periods depending on the coefficients $g_{i}$.
Nevertheless, since the $g_{i}$ are aleatory, the periods of the individual
functions are different and, as a consequence, the recurrence time of $%
r_{1}(t)$ will be very large, and strongly increasing with the number $N$ of
particles.

The time-behavior of $r_{1}(t)$ was computed by means of a numerical
simulation, where the aleatory numbers $\left\vert \alpha _{i}\right\vert
^{2}$, $\left\vert \beta _{i}\right\vert ^{2}$\ and $g_{i}$\ were obtained
from a generator of aleatory numbers: these generator fixed the value of $%
\left\vert \alpha _{i}\right\vert ^{2}$, and the $\left\vert \beta
_{i}\right\vert ^{2}$ were computed as $\left\vert \beta _{i}\right\vert
^{2}=1-\left\vert \alpha _{i}\right\vert ^{2}$. The function $r_{1}(t)$ for $%
N=200$ is plotted in Figure 1 (see also numerical simulations in 
\cite{Max}), which shows that the system $P$ decoheres in interaction with
an environment of $N$ particles $P_{i}$.

\begin{figure}[t]
\par
\centerline{\scalebox{0.7}{\epsffile{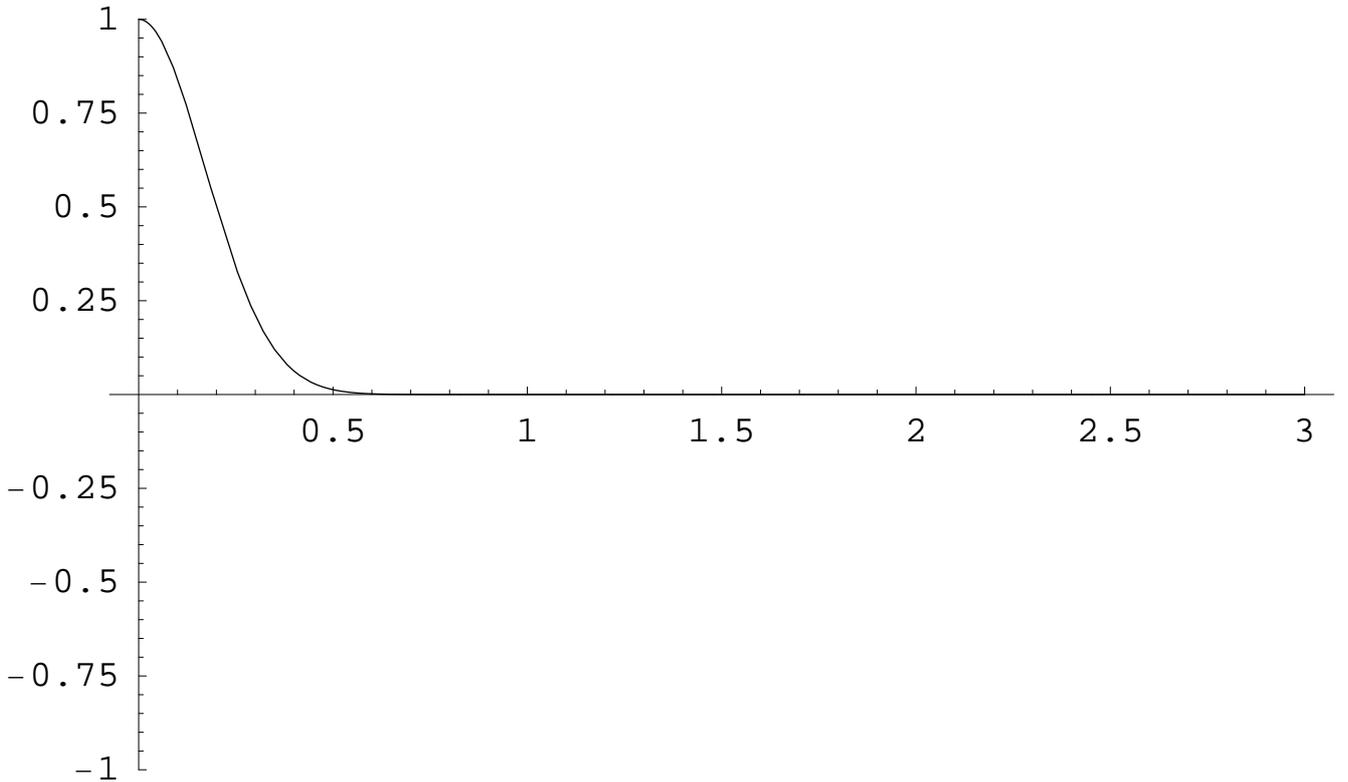}}} 
\vspace*{0.cm}
\caption{Decoherence for $S=P$ with $N=200$.}
\end{figure}

\paragraph{\textbf{Case 2: Observing the particles }$P_{i}$\textbf{.}}

Although in the usual presentations of the model the system of interest is $%
P $, as in the previous section, we can conceive different ways of splitting
the whole $U$ into an open system and an environment. For instance, it may
be the case that the measuring arrangement \textquotedblleft
observes\textquotedblright\ a subset of the particles of the environment,
e.g., the $p$ first particles $P_{j}$. \ In this case, the system of
interest is composed by $p$ particles, $S=$ $\sum\limits_{i=1}^{p}P_{i}$,
and the environment is composed by all the remaining particles, $%
E=P+\sum_{i=p+1}^{N}P_{i}$. \ So, in eq. (\ref{3.7}), $s_{\Uparrow \Uparrow
}=s_{\Downarrow \Downarrow }=1$, $s_{\Uparrow \Downarrow }=s_{\Downarrow
\Uparrow }=0$, the coefficients $\epsilon _{\uparrow \uparrow }^{(j)}$, $%
\epsilon _{\downarrow \downarrow }^{(j)}$, $\epsilon _{\downarrow \uparrow
}^{(j)}$ are generic for $j\in \left\{ 1...p\right\} $, and $\epsilon
_{\uparrow \uparrow }^{(i)}=\epsilon _{\downarrow \downarrow }^{(i)}=1$, $%
\epsilon _{\downarrow \uparrow }^{(i)}=\epsilon _{\uparrow \downarrow
}^{(i)}=0$ for $i\in \left\{ p+1...N\right\} $. Then, the relevant
observables $O_{R}\in \mathcal{O}_{R}\subset \mathcal{O}$\ read%
\begin{equation}
O_{R}=I_{S}\otimes \left( \bigotimes_{j=1}^{p}O_{S_{j}}\right) \otimes
\left( \bigotimes_{i=p+1}^{N}I_{i}\right)  \label{5-1.1}
\end{equation}%
where $O_{S_{j}}$ is given by

\begin{eqnarray}
O_{S_{j}} &=&\epsilon _{\uparrow \uparrow }^{(j)}\,|\uparrow _{j}\rangle
\langle \uparrow _{j}|+\epsilon _{\downarrow \downarrow }^{(j)}\,|\downarrow
_{j}\rangle \langle \downarrow _{j}|  \notag \\
&&+\epsilon _{\downarrow \uparrow }^{(j)}\,|\downarrow _{j}\rangle \langle
\uparrow _{j}|+\epsilon _{\uparrow \downarrow }^{(j)}\,|\uparrow _{j}\rangle
\langle \downarrow _{j}|  \label{4-1.2}
\end{eqnarray}%
Therefore, the expectation value of the relevant observables $O_{R}$ is 
\begin{equation}
\langle O_{R}\rangle _{\psi (t)}=\prod_{i=1}^{p}\left[ 
\begin{array}{c}
|\alpha _{i}|^{2}\epsilon _{\uparrow \uparrow }^{(i)}+\alpha _{i}{}^{\ast
}\beta _{i}\epsilon _{\uparrow \downarrow }^{(i)}e^{-ig_{i}t} \\ 
+|\beta _{i}|^{2}\epsilon _{\downarrow \downarrow }^{(i)}+(\alpha
_{i}{}^{\ast }\beta _{i}\epsilon _{\uparrow \downarrow }^{(i)})^{\ast
}e^{ig_{i}t}%
\end{array}%
\right]  \label{5-1.4}
\end{equation}

If $p=1$, the expectation value of eq. (\ref{5-1.4}) results

\begin{eqnarray}
\langle O_{R_{j}}\rangle _{\psi (t)} &=&\left\vert \alpha _{j}\right\vert
^{2}\epsilon _{\uparrow \uparrow }^{(j)}+\left\vert \beta _{j}\right\vert
^{2}\epsilon _{\downarrow \downarrow }^{(j)}  \notag \\
&&+\func{Re}\left( \alpha _{j}\beta _{j}^{\ast }\epsilon _{\uparrow
\downarrow }^{(j)}e^{ig_{j}t}\right)  \label{4-1.3}
\end{eqnarray}%
The evolution of $\langle O_{R_{j}}\rangle _{\psi (t)}$ depends on the
time-behavior of the third term of eq. (\ref{4-1.3}), which can rewritten as%
\begin{equation}
r_{2}(t)=\func{Re}\left( \alpha _{j}\beta _{j}^{\ast }\epsilon _{\uparrow
\downarrow }^{(j)}e^{ig_{j}t}\right)  \label{4-1.4}
\end{equation}%
In this case, numerical simulations are not required to see that $r_{2}(t)$
is an oscillating function which, as a consequence, has no limit for $%
t\rightarrow \infty $. This means that a single particle $S=P_{j}$ with a
large environment $E=P+\sum_{i\neq j}P_{i}$ of $N$ particles does not
decohere.\ Nevertheless, this result can be understood by considering that $%
P_{j}$ strongly interacts only with particle $P$, but does not interact with
the rest of the particles $P_{i\neq j}$; therefore, the interaction of $%
S=P_{j}$ with its environment $E=P+\sum_{i\neq j}P_{i}$ is not strong enough
to produce decoherence

In order to obtain the expectation value $\langle O_{R_{j}}\rangle _{\psi
(t)}$ for $p>1$, we will simplify the computation by considering the
particular case for which the relevant observables are

\begin{equation}
O_{R}=I_{S}\otimes \left( \bigotimes_{j=1}^{p}S_{x}^{(j)}\right) \otimes
\left( \bigotimes_{i=p+1}^{N}I_{i}\right)  \label{6-1.1}
\end{equation}%
where $S_{x}^{(j)}$ is the projection of the spin onto the $x$-axis of the
particle $P_{j}$. Then, $\epsilon _{\uparrow \uparrow }^{(j)}=\epsilon
_{\downarrow \downarrow }^{(j)}=0$, and the expectation value reads 
\begin{equation}
\langle O_{R}\rangle _{\psi (t)}=r_{3}(t)=\prod_{i=1}^{p}\left[ 2\func{Re}%
\left( \alpha _{i}{}^{\ast }\beta _{i}\epsilon _{\uparrow \downarrow
}^{(i)}e^{-ig_{i}t}\right) \right]  \label{6-1.2}
\end{equation}

\begin{figure}[t]
\par
\centerline{\scalebox{0.7}{\epsffile{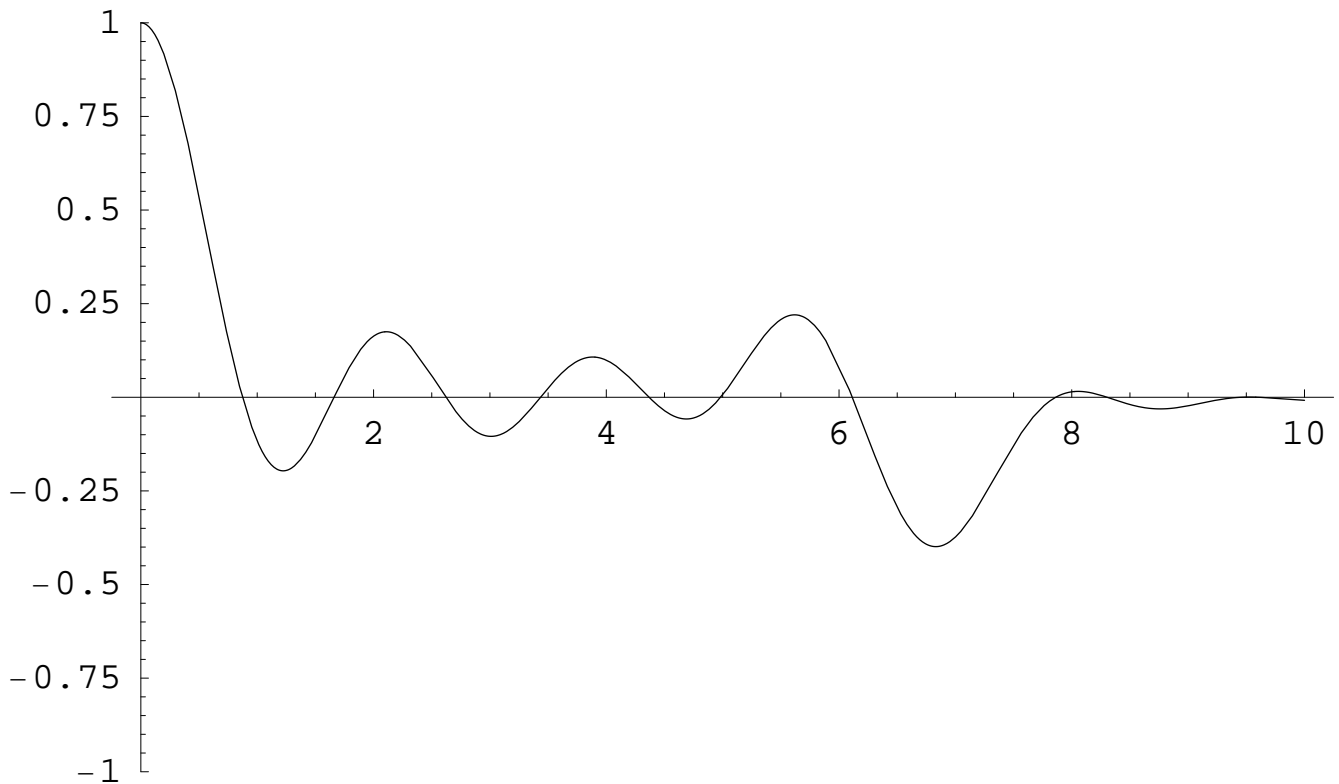}}} 
\vspace*{0.cm}
\caption{Plot of $r_{3}(t)$ given by
eq. (\protect\ref{6-1.2}), for $p=4$.}
\end{figure}

The time-behavior of $r_{3}(t)$, with $p=4$, is plotted in Figure 2, where
we can see a fast decaying followed by fluctuations around zero. As
expected, such fluctuations strongly damp off with the increase of the
number $p$ of particles, as shown in Figure 3 ($p=8$) and Figure 4 ($p=10$);
with $p=200$ the plot turns out to be indistinguishable of that obtained for
the decoherence of Case 1 with $N=200$.

\begin{figure}[t]
\par
\centerline{\scalebox{0.7}{\epsffile{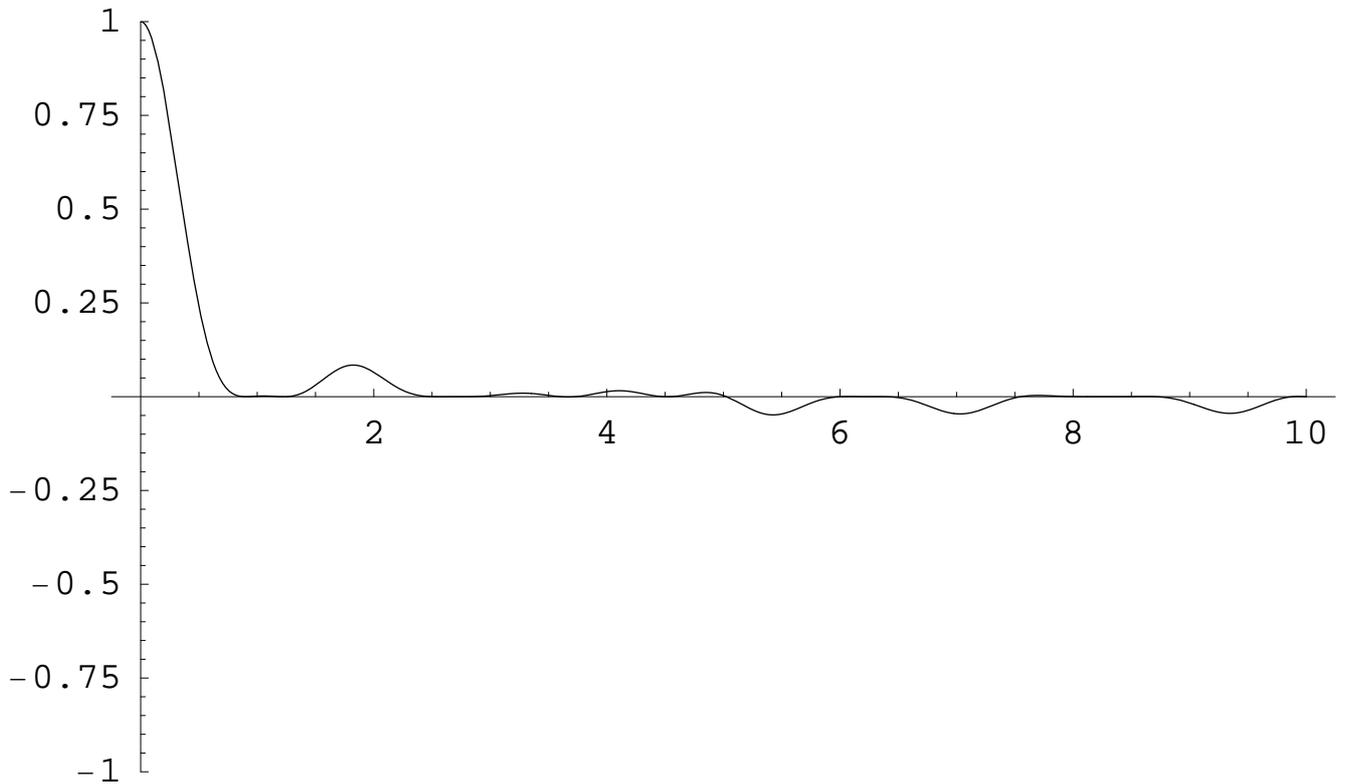}}} 
\vspace*{0.cm}
\caption{Plot of $r_{3}(t)$ given by
eq. (\protect\ref{6-1.2}), for $p=8$.}
\end{figure}

\begin{figure}[t]
\par
\centerline{\scalebox{0.7}{\epsffile{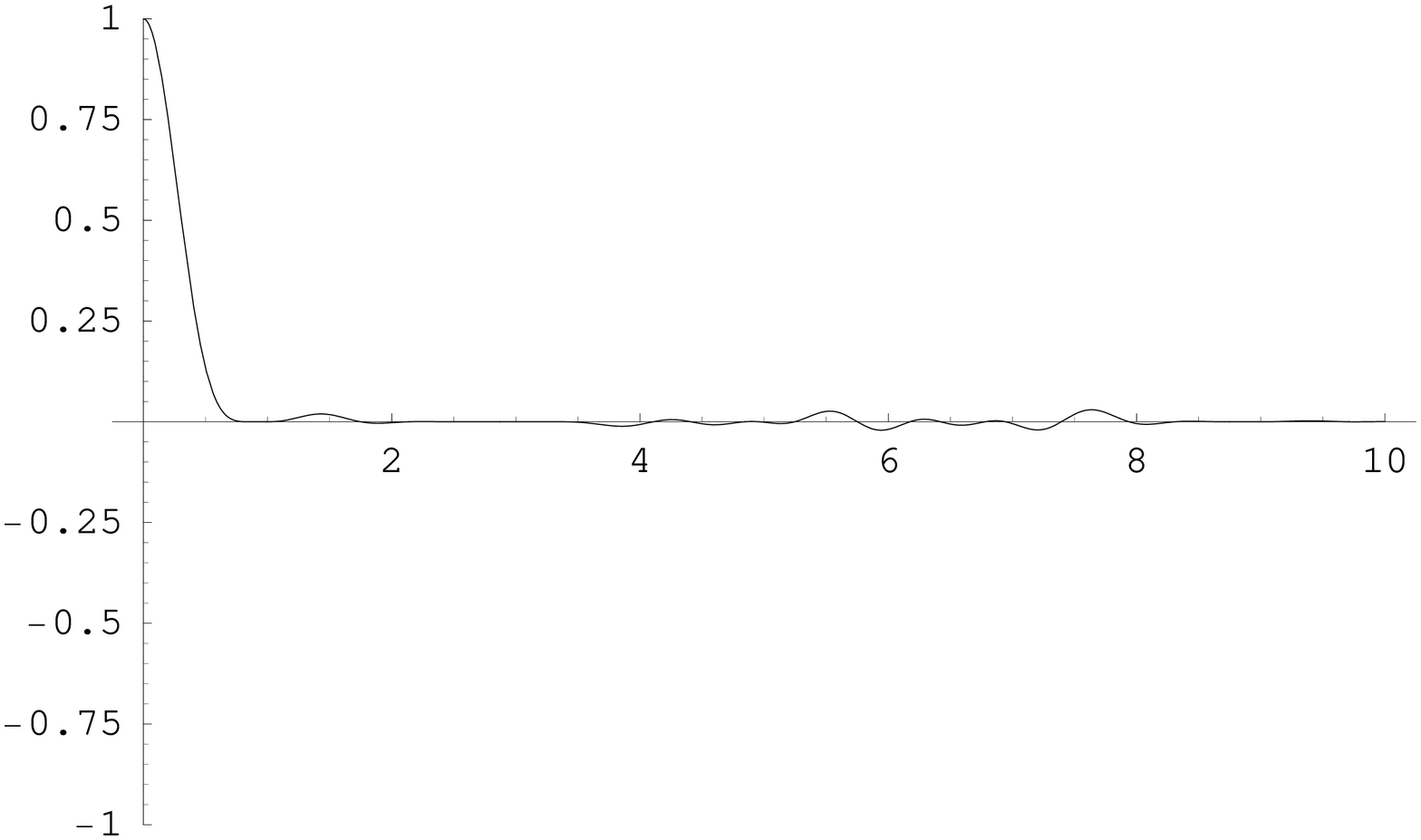}}} 
\vspace*{0.cm}
\caption{Plot of $r_{3}(t)$ given by
eq. (\protect\ref{6-1.2}), for $p=10$.}
\end{figure}

The surprising consequence of these results is that the time-behavior is
independent of the number $N$ of the particles $P_{i}$, but only depends on
the number $p$ of the particles that constitute the system of interest (see
eq. (\ref{5-1.4})). Therefore, we can consider a limit case of $N=p=10$,
where the system $S$ is composed by the $p=N=10$ particles and the
environment $E$ is a single particle, $E=P$: in this case, as shown in
Figure 4, we have to say that a system of $10$ particles decoheres as the
result of its interaction with a single-particle environment.\ The situation
becomes even more striking as the number $p$ increases: with $N=p=200$, the
system of $200$ particles strongly decoheres in interaction with a
single-particle environment.

\paragraph{\textbf{Conclusions.}}

The need of selecting a set of relevant observables, in terms of which the
time-evolution of the system is described, is explicitly or implicitly
admitted by the different approaches to the emergence of classicality: gross
observables in van Kampen \cite{van Kampen}, macroscopic observables of the
apparatus in Daneri \textit{et al.} \cite{Daneri}, collective observables in
Omnès \cite{Omnes-1994}, \cite{Omnes-1999}. \ It is quite clear that a
closed system can be \textquotedblleft partitioned\textquotedblright\ into
many different ways and, thus, there is not a single set of relevant
observables essentially privileged (see \cite{Sujeeva-1}, \cite{Sujeeva-2}).
Each partition depends on the experimental viewpoint adopted, and represents
a decision about which degrees of freedom are to be \textquotedblleft
observed\textquotedblright\ and which are disregarded in each case.\ Since
there is no privileged or essential partition, there is no need of an
unequivocal criterion to decide where to place the cut between
\textquotedblleft the\textquotedblright\ system and \textquotedblleft
the\textquotedblright\ environment: the \textquotedblleft looming
big\textquotedblright\ problem of defining the systems that decohere
vanishes when the relativity of decoherence is recognized.

This conclusion is a natural consequence of the fact that the dynamical
postulate of quantum mechanics refers to closed systems: the time-behavior
of the parts resulting from different partitions of the closed system has to
be inferred from that postulate. Since the total Hamiltonian rules the
dynamical evolution of the closed system, then the time-behavior of its
parts depends on the form in which the Hamiltonian is decomposed in each
particular partition. This means that the occurrence of decoherence cannot
be simply inferred from the interaction between a small open system and a
large environment: the decomposition of the total Hamiltonian has to be
studied in detail in each case, in order to know whether the system of
interest resulting from the partition decoheres or not under the action of
its self-Hamiltonian and the interaction Hamiltonian. As we have seen, when
the phenomenon of decoherence is studied from this perspective, certain
unexpected results are obtained, as the case of a system decohering in
interaction with a very small environment.\ Such a result disagrees with the
standard reading of the phenomenon, according to which the dissipation of
information and energy from the system to a very large environment is what
causes the destruction of the coherence between the states of the system.

\paragraph{\textbf{Acknowledgments.}}

We are very grateful to Roland Omnès and Maximilian Schlösshauer for many
comments and criticisms. This research was partially supported by grants of
the University of Buenos Aires, CONICET and FONCYT of Argentina.

\end{document}